\documentclass[aps,prx,twocolumn,preprintnumbers,superscriptaddress,10pt,showkeys]{revtex4-2}
\usepackage{graphicx}
\usepackage{dcolumn}
\usepackage{bm}
\usepackage{hyperref}
\usepackage{colortbl}
\usepackage{amsmath}
\newcommand{\dd}{\mathrm{d}}

\begin{document}

\title{Locating the QCD critical point with neutron-star observations}

\author{Christian Ecker}
\email{ecker@itp.uni-frankfurt.de}
\affiliation{Institut f\"ur Theoretische Physik, Goethe Universit\"at, Max-von-Laue-Str. 1, 60438 Frankfurt am Main, Germany}
\author{Niko Jokela}
\email{niko.jokela@helsinki.fi}
\affiliation{Helsinki Institute of Physics and Department of Physics, P.O. Box 64, FI-00014 University of Helsinki,
Finland}
\author{Matti J\"arvinen}
\email{matti.jarvinen@apctp.org}
\affiliation{Asia Pacific Center for Theoretical Physics, Pohang, 37673, Korea}
\affiliation{Department of Physics, Pohang University of Science and Technology, Pohang 37673, Republic of Korea}

\begin{abstract}
We present a probabilistic model for the QCD critical endpoint (CEP) and the equation of state (EOS) at $\beta$-equilibrium, constrained by neutron-star observations. 
Using a hybrid framework that combines the holographic V-QCD model with an effective van der Waals description of nuclear matter, we generate a large ensemble of EOSs incorporating nuclear theory uncertainties.
Constraining this ensemble with neutron-star mass-radius data and tidal deformability measurements from gravitational waves, we identify a strong first-order deconfinement transition at zero temperature, with a transition strength of $\Delta n_{\rm{PT}}/n_0 = 3.75^{+0.70}_{-0.53}$ and onset density $n_{\rm{PT}}/n_0 = 4.9^{+1.33}_{-1.17}$. The resulting posterior yields $95\%$ credible intervals for the CEP location: $\mu_{\rm{crit}} = 626^{+90}_{-179}\,\rm{MeV}$, $T_{\rm{crit}} = 119^{+14}_{-6}\,\rm{MeV}$.
\end{abstract}


\preprint{APCTP Pre2025 - 012}
\preprint{HIP-2025-18/TH}

\maketitle

\section{Introduction\label{sec:Intro}}

Understanding the phase diagram of quantum chromodynamics (QCD) is essential for describing matter under extreme conditions, from the dense cores of neutron stars (NSs)~\cite{Vuorinen:2024qws} to the high-energy environments of heavy-ion collisions (HIC)~\cite{Du:2024wjm}.
A central challenge lies in reliably capturing the transition between hadronic matter and the deconfined quark-gluon plasma, especially in regions of the phase diagram that remain inaccessible to first-principles calculations.
Recent observations from gravitational wave (GW) astronomy, particularly those involving binary NS mergers, have begun to constrain the properties of dense QCD matter, offering a unique opportunity to tighten the theoretical landscape.

In this Letter, we introduce a framework that combines a hybrid holographic equation of state (EOS) model directly with experimental and observational constraints. 
To this end, the holographic V-QCD model~\cite{Jarvinen:2011qe} with an effective van der Waals (vdW) description of nuclear matter (NM)~\cite{Demircik:2021zll} is used to construct a broad ensemble of EOSs, which is then confronted with data.
This approach~\cite{Jarvinen:2021jbd,Hoyos:2021uff} incorporates both nuclear theory uncertainties and astrophysical constraints, enabling a systematic exploration of QCD phase structure at high baryon density.

Using input from direct mass measurements of $M \gtrsim2\,M_\odot$~\cite{Cromartie:2019kug,Fonseca:2021wxt}, NICER’s mass–radius constraints~\cite{Riley:2019yda,Miller:2019cac,Riley:2021pdl}, and LIGO’s tidal deformability measurement $\tilde\Lambda\lesssim720$ from GW170817~\cite{Abbott:2018exr} we identify preferred regions in the EOS parameter space that support a strong first-order deconfinement phase transition (PT) and locate its onset at densities comparable to those reached in the most massive observed NSs~\cite{Lattimer:2021emm}.

Our analysis further yields estimates for the chemical potential and temperature of the QCD critical endpoint (CEP)~\cite{Halasz:1998qr} while remaining consistent with lattice QCD at low density~\cite{Jokela:2018ers} and cold NS matter at high density~\cite{Jokela:2021vwy,Jokela:2020piw}.
That is, the use of the holographic model allows us to construct a quantitatively constrained, nonperturbative description of QCD thermodynamics across a wide range of conditions.
Beyond its relevance to astrophysical modeling, the framework can be directly applied to HIC, where it enables testable predictions for critical phenomena.
This work establishes a novel connection between GW observations and collider experiments, providing a unified tool to study QCD matter across distinct regimes and addressing emerging priorities to link astrophysical and nuclear physics frontiers~\cite{Lovato:2022vgq}.

Being of fundamentally different nature, our holographic approach complements other studies of the QCD EOS and CEP, including applications of 
Dyson-Schwinger~\cite{Gunkel:2021oya} and functional renormalization group approaches~\cite{Fu:2019hdw,Gao:2020fbl} to QCD, as well as Nambu--Jona-Lasinio, its extensions (see, \emph{e.g.},~\cite{Buballa:2003qv,Roessner:2006xn,Costa:2008yh}), and chiral mean field models~\cite{Steinheimer:2025hsr}.
Combining future high-precision neutron-star observations with machine learning  techniques~\cite{Fujimoto:2019hxv,Carvalho:2023ele,DiClemente:2025pbl} could also allow reconstructing possible PTs in cold dense matter~\cite{Li:2025obt}.

\section{EOS hybrid construction}

The basic idea is to follow the hybrid approach of~\cite{Demircik:2021zll}, which combined nuclear theory, vdW, and holographic models into an overall state-of-the-art EOS at finite temperature and density. 
However, this construction will be generalized as follows: We allow a more general dependence on the parameters of the holographic and the vdW models.
Moreover, we replace the single nuclear theory EOS of~\cite{Demircik:2021zll} near nuclear saturation density $n_0\approx 0.16$ fm${}^{-3}$ with a sampled family.

Let us review the parameters of the hybrid construction. While the holographic V-QCD model~\cite{Bigazzi:2005md,Casero:2007ae,Gursoy:2007cb,Gursoy:2007er,Jarvinen:2011qe} in principle contains a huge number of parameters, these are strongly constrained by lattice data for the thermodynamics of QCD~\cite{Gursoy:2009jd,Panero:2009tv,Jokela:2018ers}.
As for the holographic model of the unpaired quark matter (QM) EOS, this leaves only one parameter $W_0$~\cite{Jokela:2018ers} unfixed~\footnote{Ideally, one would also sample over lattice QCD uncertainties. However, their impact is subleading compared to other error sources, and their correlated systematic nature makes consistent probabilistic treatment challenging.
}.
Including holographic NM following the approach of~\cite{Ishii:2019gta,Ecker:2019xrw,Jokela:2020piw} adds two parameters, $b$ and $c$, which control the scale of the chemical potential $\mu$ and the normalization 
of pressure $p$, respectively.
Moreover, the vdW model, which is only used to describe the temperature dependence of the NM EOS, adds the excluded volume of the nucleons, $v_0$. 
Finally, the crust EOS will be described by a linear interpolation between sampled values of the adiabatic speed of sound squared $c_s^2=\dd p/\dd e$, where $e$ denotes the energy density.

\subsection{Components of the cold EOS}

The cold EOS model consists of four parts.
For the first part we take a cold ($T=0.1~$MeV) $\beta$-equilibrium slice of the statistical Hempel--Schaffner-Bielich (HS) model~\cite{Hempel:2009mc}.
There are many implementations of this model available on the CompOSE database~\cite{Typel:2013rza,CompOSECoreTeam:2022ddl}, but for concreteness we use Steiner--Fischer--Hempel (SFHo) EOS~\cite{Moller:1996uf,Steiner:2012rk}\footnote{We model the low-density regime using the SFHo EOS instead of the commonly used Baym--Pethick--Sutherland (BPS) EOS~\cite{Baym71b} because SFHo includes a finite-temperature extension, which we adopt in the full hybrid framework. The two models agree well across their overlapping density range.}.
This EOS is used in the range $\mu\leq\mu_{\rm HS}$ of baryon number chemical potentials, where $\mu_{\rm HS}\approx 951~$MeV corresponds to a matching baryon number density $n_{\rm HS}=0.4~n_0$ and can in principle be chosen arbitrarily within reasonable bounds.

The second part of the construction is built from a continuous chain of $N$ segments with piecewise-linear $c_s^2$ as a function of $\mu$~\cite{Annala:2019puf},
\begin{equation}
    \label{eq:cs2}
    c_s^2(\mu)=\frac{\left(\mu _{i+1}-\mu \right) c_{s,i}^2+\left(\mu -\mu_i\right) c_{s,i+1}^2{}}{\mu _{i+1}-\mu _i}\,,\ \mu_i\leq\mu\leq\mu_{i+1}\,.
\end{equation}
We choose the matching point coefficients for $\mu_i$ equally separated between $\mu_0=\mu_{\rm HS}$ and $\mu_N=\mu_*$, the matching value to the V-QCD model. 
Its expected value is $1~\rm GeV$ and will be determined through a matching procedure.
The coefficients $c_{s,i}^2 \in [c_{s,\rm HS}^2,c_{s,\rm V-QCD}^2]$ are uniformly sampled between the corresponding values $c_{s,0}^2=c_{s,\rm HS}^2\approx 0.01$ and $c_{s,N}^2=c_{s,\rm VQCD}^2$ of the HS and the V-QCD model, respectively.
Furthermore, we assume monotonicity with density $c_{s,i}^2<c_{s,i+1}^2$, such as expected in pure NM.
The corresponding number density and pressure follow then from thermodynamic relations
\begin{equation}
    n(\mu)=n_{\rm HS}\,e^{\int_{\mu_{\rm HS}}^\mu \dd\mu'\frac{1}{\mu'c_s^2(\mu')}} \ , \  p(\mu)= p_{\rm HS}+\int_{\mu_{\rm HS}}^\mu \dd\mu' n(\mu')\,, \label{eq:pcrust}
\end{equation}
where the integration constants $n_{\rm HS}$ and $p_{\rm HS}\approx 0.3$~MeV/fm$^3$ are fixed by HS model at $\mu=\mu_{\rm HS}$. 

The high-density component of the cold EOS is the QM EOS from the V-QCD model, following~\cite{Jokela:2018ers}.
Since dependence on the single parameter $W_0$ is rather weak, we carry out a simple quadratic interpolation (at fixed $T$ and $\mu$) to the $W_0$ dependence of the 5b, 7a, and 8b EOSs of~\cite{Jokela:2018ers}, which have $W_0=1$, $2.5$, and $5.886$, respectively. This fit then generalizes the EOS to other values of $W_0$.

As the last component, beyond the matching point $\mu = \mu_*$, we extend the NM EOS using the baryonic sector of the V-QCD model. 
To explore a broad range of possible EOS behaviors in this regime, we generate a family of extensions by solving the V-QCD model for various values of its two parameters, $W_0$ and $b$, within their allowed ranges. 
The parameter $c$ only affects the normalization and will be added in the matching procedure.
We define a new parameter $\delta b$ which gives the relative deviation of $b$ from a typical value, given by $b=9.4$, $10.5$, and $31.4$ for EOSs 5b, 7a, and 8b, respectively~\cite{Jokela:2018ers,Jokela:2020piw}.
The EOS depending on $W_0$ and $\delta b$ is then obtained through a similar interpolation as for the QM model.

\subsection{Matching the components}

We present the matching procedure between different components of the cold hybrid EOS. 
This is mostly simple: the crust EOS of~\eqref{eq:pcrust} is by construction continuously matched with the low density HS EOS, and the first-order PT between holographic nuclear and QM is found by searching for the point where the pressures as a function of the chemical potential match.
This will be done directly for the temperature dependent EOS as we discuss below.
However matching the two descriptions of the NM region, {\emph{i.e.}}, the piecewise-linear $c_s^2$ description and V-QCD NM, needs to be done carefully to ensure thermodynamic consistency  at $\mu = \mu_*$~\cite{Ecker:2019xrw,Jokela:2020piw}.
This is done by implementing an iterative procedure.

First, we randomly select a target baryon density $n_*\in [1.2, 2]\,n_0$, and $W_0\in [0, 6]$.
These ranges are provisional and may be refined based on empirical constraints.
We also initially fix $c = 3$ and $\delta b = 0$.
We determine the initial $\mu_*$ by requiring that $n_{\rm V-QCD}(\mu_*) = n_*$. 

The iterative process continues as follows, with the order of operations optimized for convergence.
Using the SFHo nuclear EOS as the baseline, we construct a sampled piecewise extension of the crust EOS such that speed of sound matches the V-QCD prediction
at the chemical potential $\mu_*$, {\emph{i.e.}}, we enforce $c_s^2(\mu_*) = c_s^2(\mu_*)_{\rm V-QCD}$ with the V-QCD result evaluated at the initially chosen $c$ and $\delta b$.
We then scale $\mu$ linearly, keeping the lower endpoint $\mu_{\rm HS}$ of the sampled region fixed, and consequently varying $\mu_*$ so that the baryon number density of the crust EOS matches with $n_*$ at the new $\mu_*$. 

At this point neither $p$ nor $n$ (due to the scaling of $\mu_*$) match between the crust and V-QCD EOSs.
To continue, we solve for $\delta b$ by requiring continuity of $p/n$ across $\mu_*$.
Once $\delta b$ is determined, we solve for the normalization parameter $c$ to maintain continuity of $n$.
Therefore the resulting EOS has continuous $p$ and $n$ at $\mu_*$ while $c_s$ is slightly discontinuous. 
This procedure iterates, returning to the step where the crust EOS is adjusted, until a suitable convergence criterion is met, though typically a single iteration is sufficient. 
In the iteration process, we do not resample the values of $c_s^2$ for the crust EOS, but first linearly rescale $c_s^{-2}$ in the crust region (with fixed $\mu_{\rm HS}$) so that the new $c_s^2$ matches with the updated V-QCD value at $\mu_*$.

\subsection{Finite temperature extension}

We construct the EOS and the PT line at finite temperature following~\cite{Demircik:2021zll}, focusing here on $\beta$-equilibrium and densities $n \gtrsim n_0$. For details at low density and beyond $\beta$-equilibrium, see~\cite{Chesler:2019osn,Demircik:2021zll}. 

First, we compute the excluded-volume corrected vdW free energy $f_{\rm ex}$.
The dependence of the extension on the excluded volume $v_0$ is rather weak. 
Therefore, it is 
enough to sample three representative values: $0.56$~fm$^{3}$, $0.8$~fm$^{3}$, and $1$~fm$^{3}$~\cite{Demircik:2021zll}.
The full NM EOS is given by
\begin{equation}
    f_\mathrm{vdW}(T,n) = f_\mathrm{cold}(n) + \Delta f_\mathrm{ex}(T,n)\ ,
\end{equation}
where $f_\mathrm{cold}(n)$ is the free energy of cold NM constructed above, and the thermal correction~\footnote{Due to the mean-field character of the potential term in the vdW interactions, it enters additively and is understood to be included in the difference between $f_\mathrm{ex}$ and $f_\mathrm{cold}$.}
\begin{align}
    &\Delta f_\mathrm{ex}(T,n) =  \underset{Y_q}{\text{min}}\big[f_\mathrm{ex}(T,n,Y_q) - f_\mathrm{ex}(0,n,Y_q) \nonumber \\
    & +  f_\mathrm{SFHo}(0,n,Y_q)\big] - f_\mathrm{SFHo}^\mathrm{\beta-eq}(0,n) + f_\mathrm{mesons}(T)
\end{align}
can be computed separately for each value of $v_0$. 
Here $f_\mathrm{SFHo}(T,n,Y_q)$ is the SFHo free energy, $f_\mathrm{SFHo}^\mathrm{\beta-eq}(T,n)$ is its value at $\beta$-equilibrium, and $Y_q$ is the proton (charge) fraction. Moreover, $f_\mathrm{mesons}(T)$ is the free energy of a noninteracting meson gas including mesons with masses up to $\sim 1~\rm GeV$ from the particle data group listings~\cite{Demircik:2021zll}.
The SFHo EOS was included as in~\cite{Demircik:2021zll}, since the vdW EOS yields an unrealistic $Y_q$-dependence, particularly underestimating the symmetry energy. 
However, at $\beta$-equilibrium the effect of this term is small.  

For the QM phase at $\beta$-equilibrium, $Y_q\approx 0$, so we directly use the V-QCD results from~\cite{Jokela:2018ers}. 
Free electron and photon gases are also included, though their contributions are negligible.

In the final step, we match the NM and QM EOSs by solving for the PT curves. 
Since we remain in $\beta$-equilibrium throughout, the PT densities $n_\mathrm{NM}^*(T)$ and $n_\mathrm{QM}^*(T)$ are determined by solving the coupled equations $\mu_\mathrm{NM}(T,n_\mathrm{NM}^*)= \mu_\mathrm{QM}(T,n_\mathrm{QM}^*)$ and $p_\mathrm{NM}(T,n_\mathrm{NM}^*) =  p_\mathrm{QM}(T,n_\mathrm{QM}^*)$ independently for each temperature.
Once the PT curves are known, they can be used to construct the total free energy $f(T,n)$.

To locate the CEP systematically, one may evaluate the latent heat $\Delta \epsilon$ across the PT as a function of temperature and determine where it vanishes. 
Following~\cite{Demircik:2021zll}, we do this by using a polynomial fit in the regime of low temperatures $T \lesssim 110$~MeV, where the latent heat is sizable. 

\section{Constraints}

To quantify the probability distribution of EOS models and infer the location of the CEP, we evaluate the posterior distribution using a product of independent likelihoods from nuclear theory and astrophysical data:
\begin{align}\label{eq:posterior}
    P(\text{EOS}|\text{data}) &\propto P(\text{CET}|\text{EOS}) P(\text{Mass}|\text{EOS}) \notag \\
    &\quad \times P(\text{NICER}|\text{EOS}) P(\tilde{\Lambda}|\text{EOS}) \ .
\end{align}
We assume a uniform prior and normalize the posterior such that the optimal model having the highest likelihood has unit weight.

As a nuclear physics constraint, we integrate each EOS up to $n \approx 2\,n_0$ using a Gaussian likelihood derived from the next-to-next-to-leading order chiral effective theory (CET) calculation of Ref.~\cite{Drischler:2020fvz}.
Our selection and implementation of astrophysical constraints closely follow Ref.~\cite{Gorda:2022jvk}.
We include mass measurement of PSR J1614-2230 ($M = 1.928 \pm 0.017~M_\odot$)~\cite{Demorest:2010bx}
obtained via pulsar timing, the simultaneous X-ray measurement of mass and radius for PSR J0740+6620 by NICER~\cite{Miller:2021qha} ($M = 2.08 \pm 0.07~M_\odot$, $R = 12.39^{+1.30}_{-0.98}~\text{km}$), and the tidal deformability constraint from the GW170817 event observed by LIGO/Virgo $\tilde{\Lambda} = 300^{+420}_{-230}$~\cite{Abbott:2018exr}.

Unlike, {\emph{e.g.}}, the generic EOS construction in~\cite{Gorda:2022jvk} we do not impose the previously used direct mass measurement of PSR J0348+0432 ($M = 2.01 \pm 0.04~M_\odot$)~\cite{Antoniadis:2013pzd}, as it has recently been revised downward to approximately $1.8~M_\odot$~\cite{Saffer:2024tlb}.
This updated value is already encompassed by the mass constraints derived from PSR J1614–2230 and PSR J0740+6620. 
Furthermore, we do not impose the perturbative QCD (pQCD) constraint explicitly, since our prior for the V-QCD QM phase naturally falls within the pQCD uncertainty band~\cite{Fraga2014,Gorda:2021znl,Gorda:2023mkk,Gorda:2023usm}\footnote{While the precise pQCD criterion of~\cite{Komoltsev:2021jzg} is generally not satisfied at high densities, our model remains consistent with it at lower densities. 
At densities above $\sim 17n_0$, this criterion would begin to constrain the parameter $W_0$ from below. 
However, at such high densities, the QCD coupling becomes small ($\alpha_s \sim 0.4$)~\cite{Kurkela:2009gj}, diminishing our confidence in the holographic approach.
Therefore, comparisons with pQCD in this regime are not meaningful.}.

\section{Results}

We present results for the EOS, NS properties, and the phase diagram.
Central panels of each figure show the normalized posterior (\ref{eq:posterior}); side panels display weighted histograms with uncertainty estimates for key quantities.

\begin{figure}[htb]
    \centering
    \includegraphics[width=0.5\textwidth]{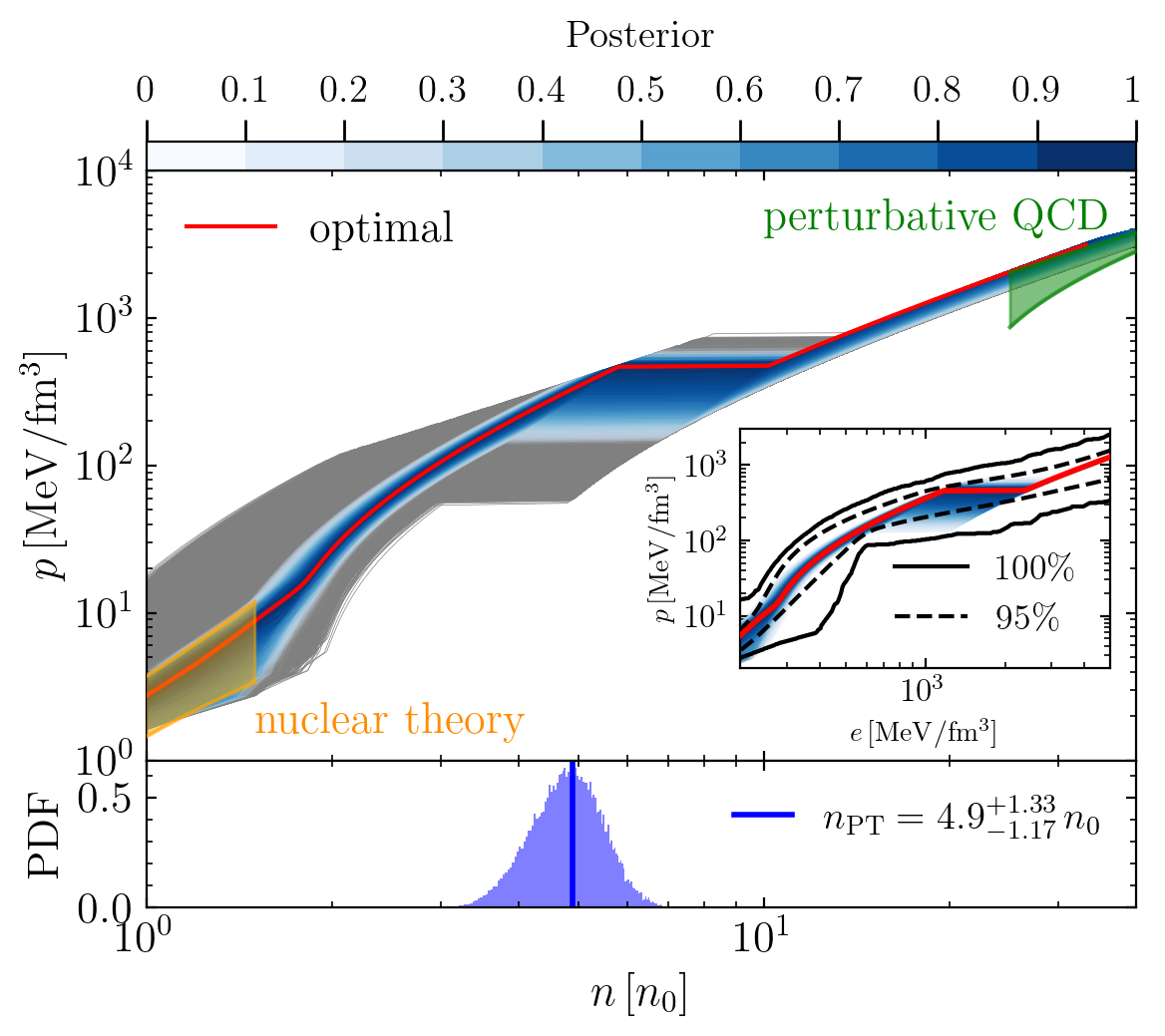}
    \caption{
        Top: Posterior probability and prior (gray) distribution for the EOS together with error estimates from nuclear theory (orange) and pQCD (green).
        Inset shows comparison to $100\%$ (solid) and $95\%$ (dashed) CIs from~\cite{Ecker:2022dlg}. 
        Bottom: Histogram of $n_{\rm PT}$.
    }
\label{fig:EOS}
\end{figure}

Fig.~\ref{fig:EOS} shows the posterior probability (PP) distribution of the zero-temperature EOS based on the prior in gray, with the highest probability model in red (repeated in all figures), and uncertainty bands from CET and pQCD in orange and green, respectively. 
The inset compares our results to a model-independent parametrization~\cite{Ecker:2022dlg} with $M_{\rm TOV} > 2.18~M_\odot$, showing that our approach yields a much narrower distribution, with the high-PP region and the optimal model well within the corresponding $95\%$ credible interval (CI).
We show histograms of the PT onset density, with median values and $95\%$ CIs given by
\begin{equation}
    n_{\rm PT} = 4.9^{+1.33}_{-1.17}\,n_0\ , \ 
    e_{\rm PT} = 935^{+335}_{-275}\,\rm MeV/fm^3\,.
\end{equation}
The sizable onset densities are a prediction of the V-QCD framework when informed by observation of NSs with $M\approx 2M_\odot$.

\begin{figure}[htb]
    \centering
    \includegraphics[width=0.5\textwidth]{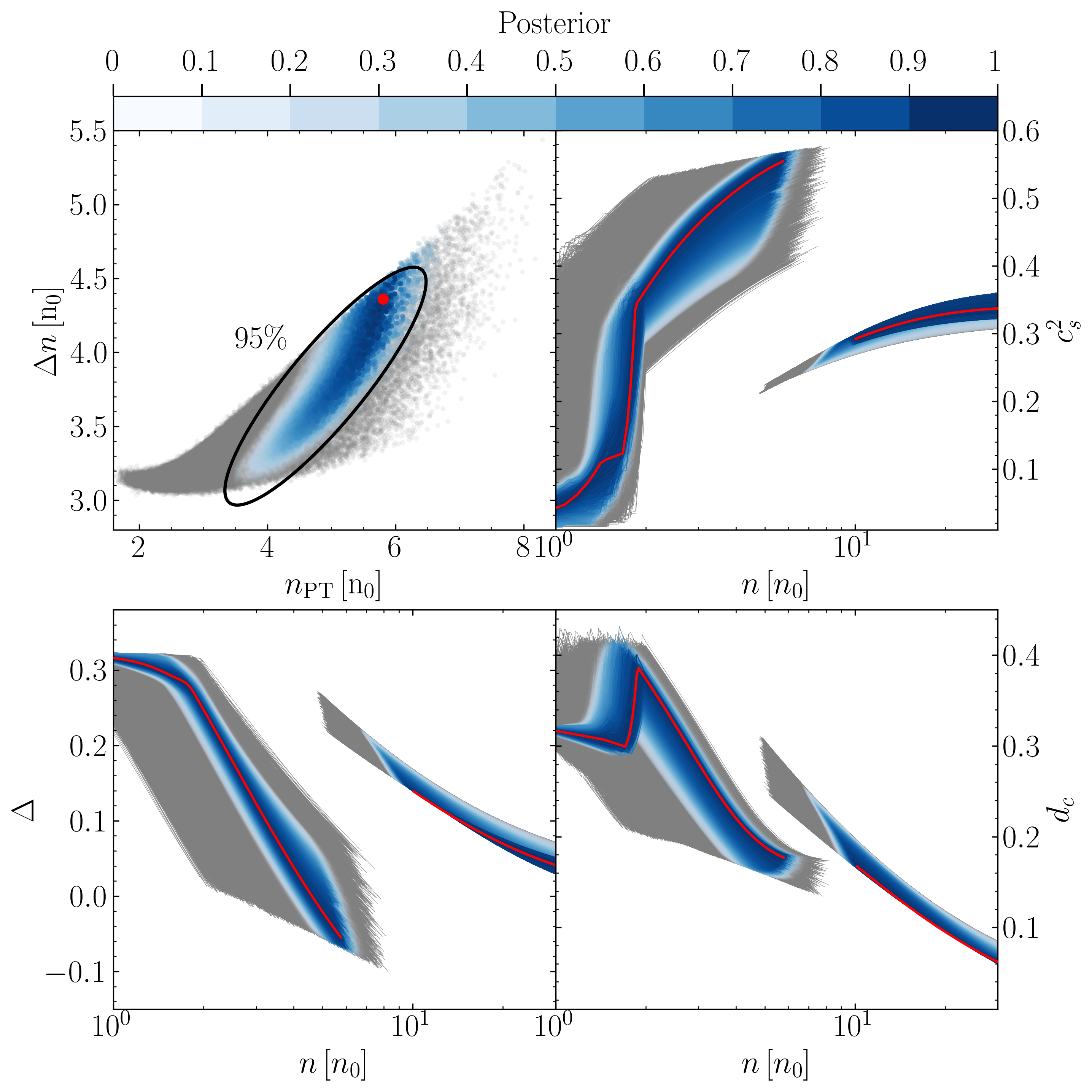}
    \caption{
        Top left: Posterior distribution of the correlation between the PT onset density ($n_{\rm PT}$) and the strength ($\Delta n$), with the $95\%$ CI shown as a black contour and the optimal model marked by a red dot.
        Top right: Distribution of $c_s^2$.
        Bottom left: Distribution of $\Delta$.
        Bottom right: Distribution of $d_c$.
    }
\label{fig:EOSproperties}
\end{figure}

Fig.~\ref{fig:EOSproperties} is a summary of cold EOS properties.
The top-left panel is the distribution for PT strength $\Delta n$ as function of the onset density for which we find
\begin{equation}
    \Delta n_{\rm PT}=3.75^{+0.70}_{-0.53}\,n_0\,.
\end{equation}
The distribution shows a strong correlation between the PT strength and the onset density, as indicated by the $95\%$ CI (black curve) of the distribution, with a preference for a strong PT consistent with the optimal model $n_{\rm PT}^{\rm opt}=5.81~n_0$, $\Delta n_{\rm PT}^{\rm opt}=4.36~n_0$.

The top-right panel illustrates the distribution of $c_s^2$.
Below $n_0$, we see a sharp rise from $c_s^2 \ll 1$ to values exceeding the conformal value ($c_s^2 > 1/3$) just before the PT density, yet remaining below $c_s^2 \lesssim 0.6$, in agreement with model-independent constructions~\cite{Altiparmak:2022,Fujimoto:2024cyv}.
The optimal model (red) peaks at $\max c_{s,\rm opt}^2=0.55$.
A gap in the distribution indicates the PT region. 
At the onset density in NM, the sound speed is higher than in the QM phase, where it shows weak density dependence and approaches the conformal limit.

The bottom-left panel shows the conformal anomaly $\Delta=1/3-p/e$~\cite{Fujimoto:2022ohj,Marczenko:2022jhl,Ecker:2022dlg}, an indicator of deviation from purely conformal behavior which is constrained by causality and thermodynamic stability to lie within the bounds $-\frac{2}{3} \leq \Delta \leq \frac{1}{3}$.
In NM, the conformal anomaly saturates its upper bound at low densities, and decreases monotonically to near-conformality ($\Delta \approx 0$) at the onset density. Throughout, $\Delta>0.1$ well above its lower bound. In QM, $\Delta$ decreases monotonically from values around $\Delta \approx 0.2$ toward zero at asymptotically large densities.

Finally, the bottom-right panel displays a related conformality measure~\cite{Annala:2023cwx,Musolino:2023edi}: $d_c=\sqrt{\Delta^2+(\Delta')^2}$, $\Delta'=\dd \Delta/\dd \log e$.
At low densities, $d_c$ approaches the value predicted by CET~\cite{Tews:2012fj,Hebeler:2013nza}, $d_c \approx 1/3$. 
It then rises rapidly, driven by a large value of $\Delta'$, to a local maximum of $d_c \approx 0.4$ at $n \approx 2\,n_0$, before decreasing monotonically to a minimum of $d_c \approx 0.17$ within the baryonic phase at the first-order PT.
At a discontinuous first-order PT, where $c_s^2 = 0$ and $\Delta' = \frac{1}{3} - \Delta$, the quantity $d_c$ is bounded from below by $\frac{1}{3\sqrt{2}} \approx 0.236$.
Ref.~\cite{Annala:2023cwx} suggests that $d_c > 0.2$ indicates the absence of QM, consistent with a purely hadronic core.
While we find good agreement with this criterion sufficiently far from the PT, it is slightly violated in its immediate vicinity: our results include NM phases with $d_c \approx 0.16$ and QM phases reaching up to $d_c \approx 0.24$.

\begin{figure}[htb]
    \centering
    \includegraphics[width=0.49\textwidth]{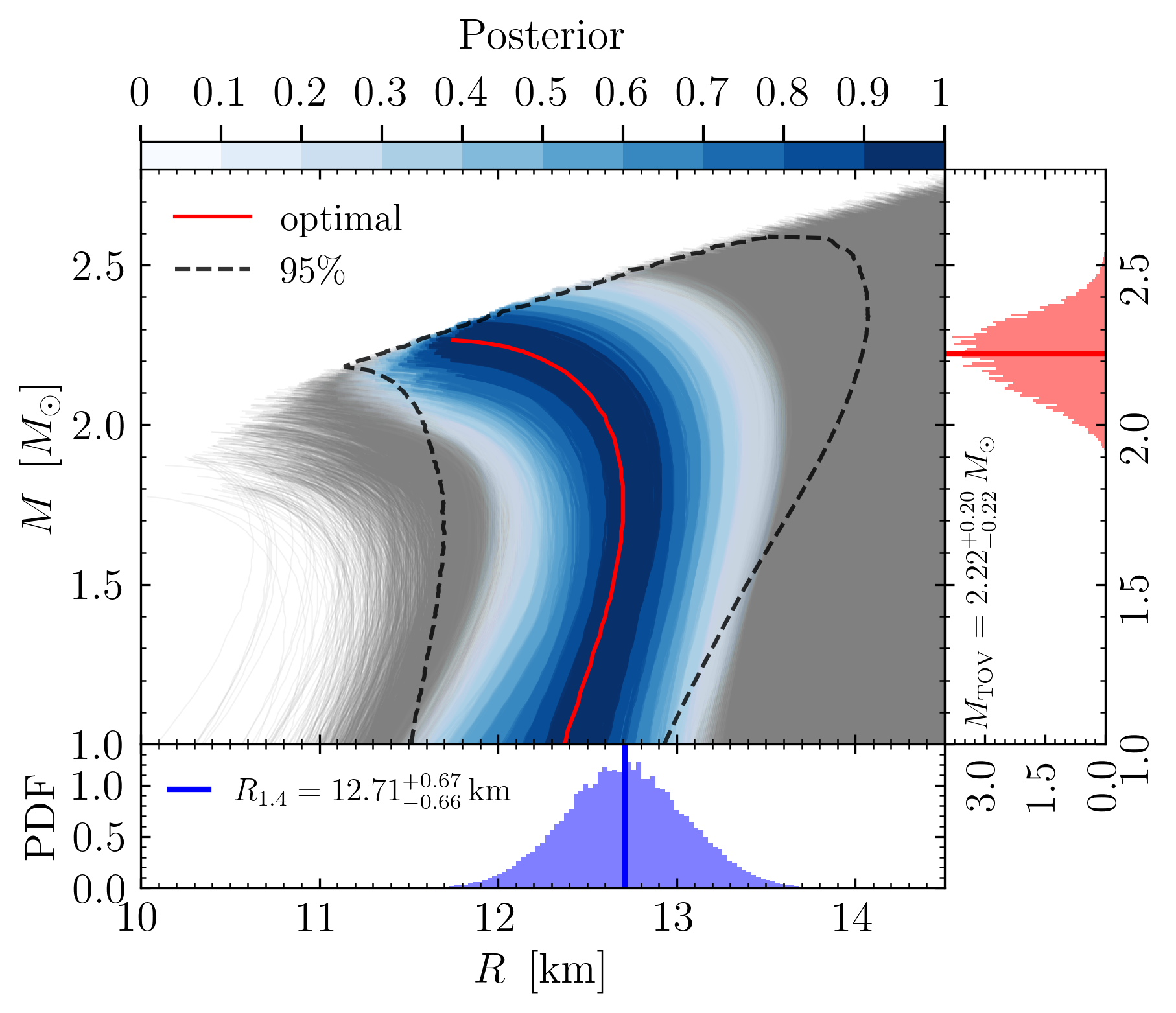}
    \caption{
        Top: Posterior distribution of NS mass–radius curves. The black dashed contour marks the $95\%$ CI from~\cite{Ecker:2022dlg}.
        Bottom: Histogram of $R_{1.4}$.
        Right: Histogram of $M_{\rm TOV}$.
    }
\label{fig:MR}
\end{figure}

Next, we present the resulting NS mass–radius relation, obtained by solving the Tolman--Oppenheimer--Volkoff (TOV) equations, as shown in Fig.~\ref{fig:MR}.
The distribution lies well within the $95\%$ (dashed) CI derived from the generic EOS ensemble~\cite{Ecker:2022dlg}, as indicated in the inset of Fig.~\ref{fig:EOS}.
Our distribution shows additional exclusion relative to~\cite{Ecker:2022dlg}, particularly in the regions of intermediate masses ($M \approx 1.5\,M_\odot$) with small radii ($R \lesssim 11.7\,\mathrm{km}$), and at large masses ($M \approx 2\,M_\odot$) with large radii ($R \gtrsim 13.5\,\mathrm{km}$).
The histogram on the bottom shows the radius distribution for a typical NS with mass $M = 1.4\,M_\odot$. 
We find
\begin{equation}
    R_{1.4}=12.71^{+0.67}_{-0.66}\,{\rm{km}}\ , \ R^{\rm opt}_{1.4}=12.6~{\rm km} \ .
\end{equation}
These are in excellent agreement with $R_{1.4}=12.72^{+0.54}_{-1.13}\,\rm{km}$ obtained in~\cite{Ecker:2022dlg}.
The histogram on the right shows the maximum mass distribution. 
We find
\begin{equation}
    M_{\rm TOV}=2.22^{+0.20}_{-0.22}\,M_\odot\ , \ M^{\rm opt}_{\rm TOV}=2.27~M_\odot \ .
\end{equation}

\begin{figure}[htb]
    \centering
    \includegraphics[width=0.49\textwidth]{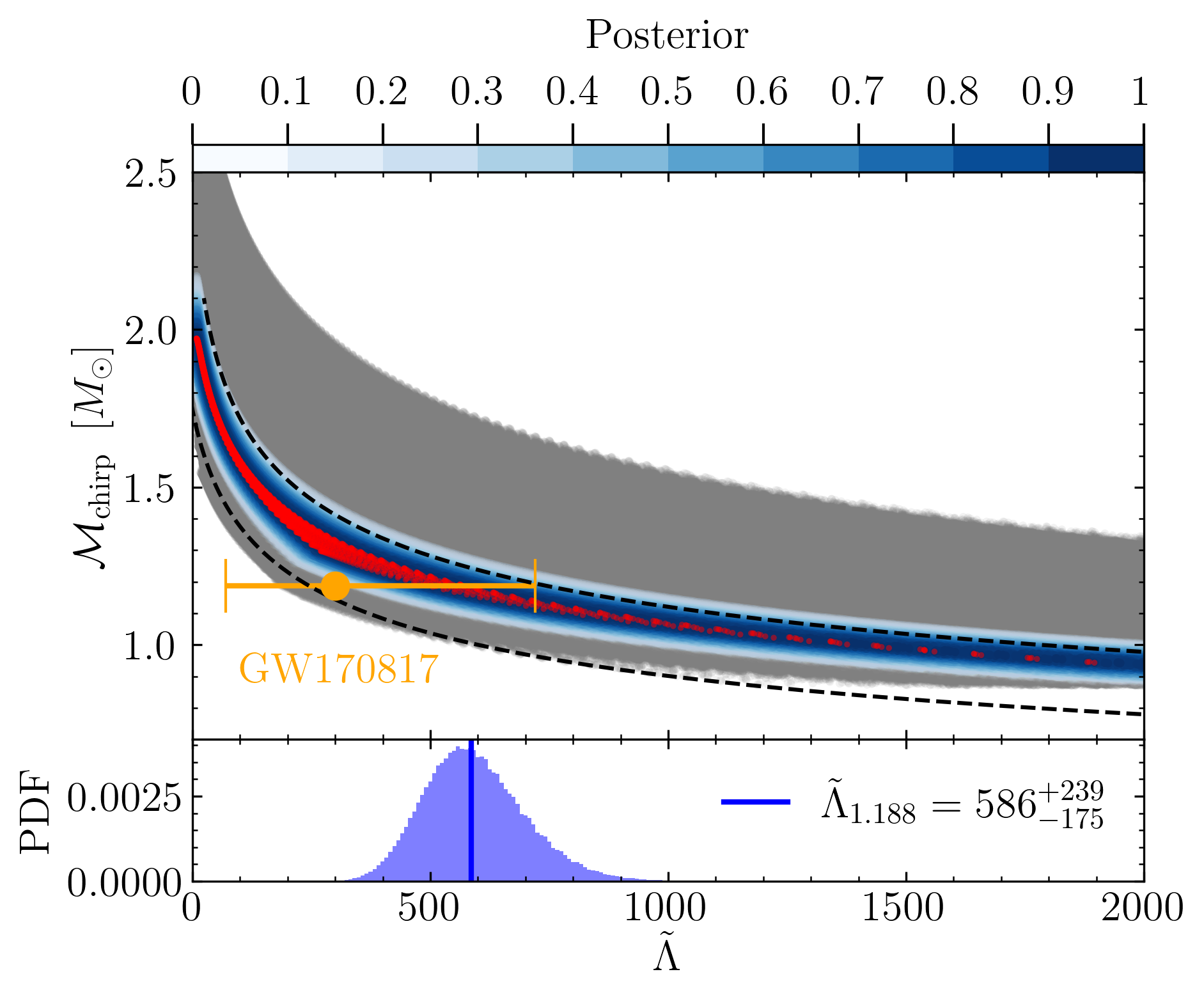}
    \caption{
        Top: Posterior distribution of the correlation between $\mathcal{M}_{\rm chirp}$ and $\tilde{\Lambda}$.
        The red point cloud shows the optimal model marginalized over mass ratios.
        The orange error bar shows the estimate~\cite{Abbott:2018exr}, and black dashed lines indicate the $95\%$ CI from Ref.~\cite{Ecker:2022dlg}. 
        Bottom: Histogram of $\tilde{\Lambda}_{1.186}$.
    }
    \label{fig:McL}
\end{figure}

Fig.~\ref{fig:McL} shows the distribution for the binary tidal deformability parameter $\tilde{\Lambda}$ and the chirp mass $\mathcal{M}_{\rm chirp}$.
This relation is particularly important for deriving EOS constraints from pre-merger GW signals of NS binaries, which determine $\tilde{\Lambda}$ at a precisely measurable $\mathcal{M}_{\rm chirp}$.
In the case of GW170817, $\mathcal{M}_{\rm chirp} = 1.186^{+0.002}_{-0.004}\,M_\odot$~\cite{LIGOScientific:2017zic}, and $\tilde\Lambda_{1.186} = 300^{+420}_{-230}$~\cite{Abbott:2018exr}, as indicated by the orange error bar.
Consistent with previous model-independent studies~\cite{Altiparmak:2022,Ecker:2022dlg} we find a strong correlation between $\tilde{\Lambda}$ and $\mathcal{M}_{\rm chirp}$, along with a slightly narrower distribution compared to generic studies, as indicated by the $95\%$ CI from~\cite{Ecker:2022dlg} (dashed lines).
From the histogram in the bottom, we extract the median value 
\begin{equation}\label{eq:Lambda}
    \tilde \Lambda_{1.186}=586^{+239}_{-175} \ .
\end{equation}
As in earlier holographic~\cite{Jokela:2020piw} and model-agnostic~\cite{Magnall:2025zhm} studies, our approach increases the inferred lower bound on $\tilde\Lambda_{1.186}$ compared to the original LIGO–Virgo result (orange error bar).
The median \eqref{eq:Lambda} is substantially larger than that of~\cite{Abbott:2018exr} and the value from physics-informed priors ($\tilde\Lambda_{1.186}=384^{+306}_{-158}$)~\cite{Magnall:2025zhm}, though with reduced uncertainty.
This stems from the stiff EOS in V-QCD NM, which yields large radii and tidal deformabilities.

\begin{figure}[htb]
    \centering
    \includegraphics[width=0.5\textwidth]{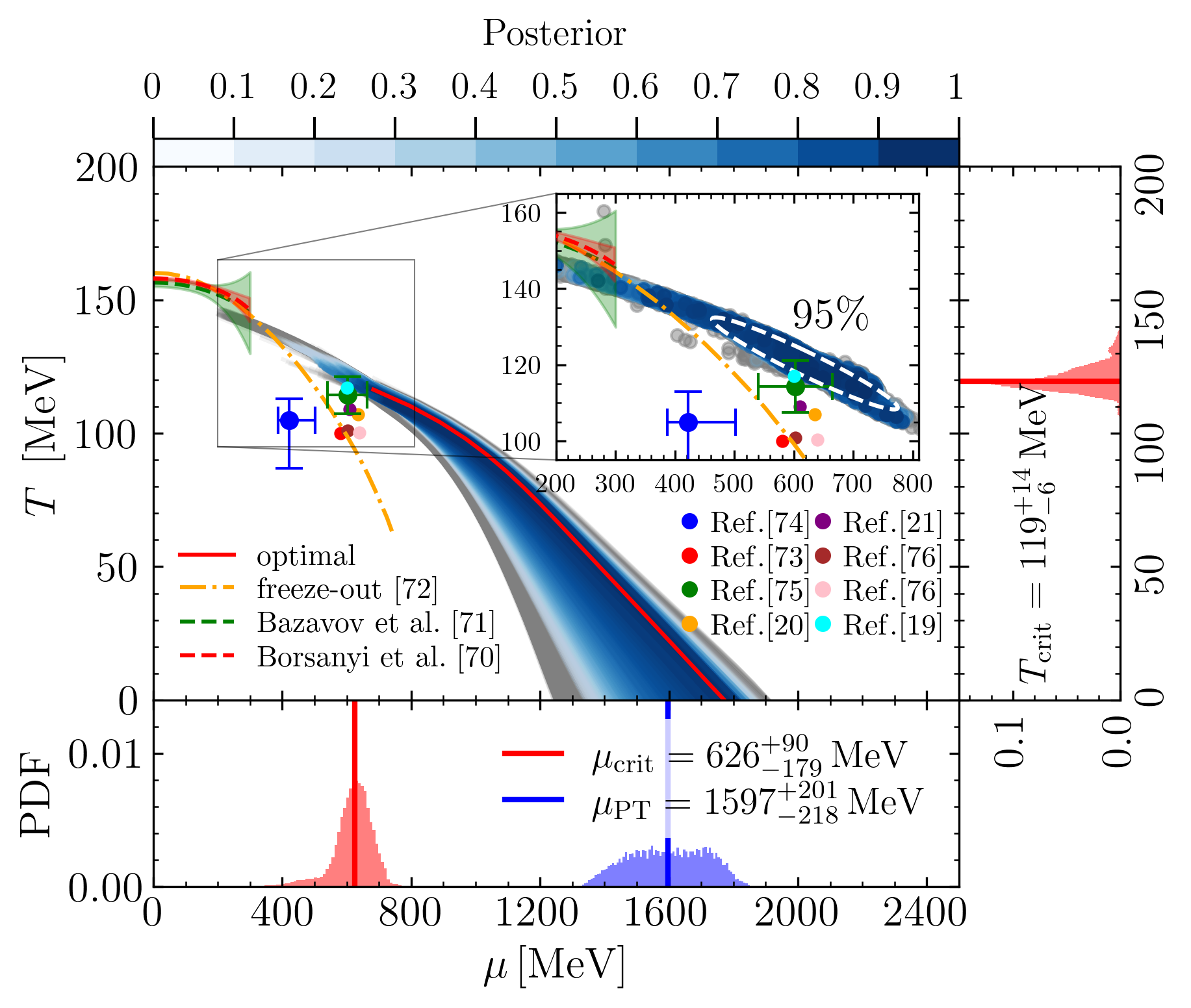}
    \caption{
        Top: Phase diagram for locally charge neutral $\beta$-equilibrated QCD matter with inset displaying the CEP.
        We show the lattice crossover lines~\cite{HotQCD:2018pds,Borsanyi:2020fev}, the chemical freeze-out line~\cite{Lysenko:2024hqp}, and small-density CEP estimates~\cite{Basar:2023nkp,Clarke:2024ugt,Shah:2024img,Fu:2019hdw,Gao:2020fbl,Gunkel:2021oya,Jokela:2024xgz}. 
        Bottom: Histograms of $\mu_{\rm crit}$ and $\mu_{\rm PT}$. 
        Right: Histogram of $T_{\rm crit}$.
    }
    \label{fig:PhaseDiagram}
\end{figure}

Finally, our main result is shown in Fig.~\ref{fig:PhaseDiagram}, displaying PT lines in the ($\mu_b$,$T$)-plane.
The inset highlights the expected critical region, showing our posterior distribution for the CEP along with the $95\%$ CI.
The results indicate a strong, approximately linear correlation between $T_{\rm crit}$ and $\mu_{\rm crit}$.
For comparison, we include recent estimates at vanishing charge and strangeness chemical potentials ($\mu_Q=0=\mu_S$) based on analytic continuation~\cite{Borsanyi:2020fev}, Taylor expansion~\cite{HotQCD:2018pds}, freeze-out analysis in the ideal HRG model~\cite{Lysenko:2024hqp}, lattice-inspired extensions~\cite{Basar:2023nkp,Clarke:2024ugt,Shah:2024img}, effective QCD models~\cite{Fu:2019hdw,Gao:2020fbl,Gunkel:2021oya}, and
black hole engineering~\cite{Jokela:2024xgz}.  

From the bottom and right panels we can infer the median values for the location of the CEP:
\begin{equation}\label{eq:critvalues}
    (\mu_{\rm crit},T_{\rm crit})=(626^{+90}_{-179},119^{+14}_{-6})\,\rm MeV\,. 
\end{equation}
The corresponding values of the optimal model are $(\mu_{\rm crit}^{\rm opt},T_{\rm crit}^{\rm opt})=(683,116)~\rm MeV$.
In addition we determine the median and optimal value of the PT chemical potential at zero temperature 
\begin{equation}\label{eq:muPT}
    \mu_{\rm PT}=1597^{+201}_{-218}\,\rm MeV\ , \ \mu_{\rm PT}^{\rm opt}=1772~\rm MeV \ .
\end{equation}

\section{Conclusion}

We analyzed the QCD phase diagram by using a unified holographic framework that is consistent with lattice QCD at small chemical potential and constrained by astrophysical observations at high baryon density.
Our framework bridges two traditionally disconnected regimes, capturing both confined and deconfined matter.
Apart from the predictions for the cold QCD EOS and neutron star properties, it extends to nonzero temperature, where 
it predicts a first-order deconfinement PT line ending at a critical point.  

Although our analysis is grounded on neutron-star matter, the same model, without further tuning, remains consistent with conditions relevant to HICs, where matter is hot, carries net charge, and is out of $\beta$-equilibrium.
The model can be straightforwardly extended to study isospin-asymmetric matter~\cite{Kovensky:2021ddl,Bartolini:2022gdf,Bartolini:2025sag}, relevant for both NSs and HICs with neutron-rich beams, helping to probe how the phase structure shifts with varying charge-to-baryon ratio.
The predicted location of the CEP can still inform the ongoing analysis of Beam Energy Scan II data.
Additionally, it provides a nonperturbative, thermodynamically consistent benchmark for future experiments at NA61/SHINE, FAIR, and NICA, potentially guiding their experimental designs and interpretations.
Moreover, the computations of conserved-charge susceptibilities~\cite{Jokela:2024xgz}, can help interpret fluctuation measurements in HIC (\emph{e.g.}, net proton or net baryon number cumulants) which are sensitive to critical phenomena. 
In addition, the optimized model we provide can serve as a realistic input for simulations of supernova modeling codes or especially for quark-matter core formation in NS merger codes~\cite{Ecker:2024kzs,Tootle:2022pvd}, a phase for which also the transport properties are known~\cite{Hoyos:2020hmq,Hoyos:2021njg,CruzRojas:2024etx}.
Finally, our predictions for the CEP may be influenced by a modulated instability~\cite{CruzRojas:2024igr,Demircik:2024aig}, which warrants further investigation.\\

\section*{Acknowlegments}
We thank Kenji Fukushima, Tyler Gorda, Fabian Rennecke, Tobias Rindlisbacher, and Aleksi Vuorinen for valuable comments, Sofia Blomqvist for her assistance with implementing various constraint data, and Christian Drischler and Tyler Gorda for providing the CET likelihood function. C.~E. acknowledges support by the Deutsche Forschungsgemeinschaft (DFG, German Research Foundation) through the CRC-TR 211 'Strong-interaction matter under extreme conditions'-- project number 315477589 -- TRR 211.
N.~J. has been supported in part by the Research Council of Finland grant no.~3545331. M.~J. has been supported by an appointment to the JRG Program at the APCTP through the Science and Technology Promotion Fund and Lottery Fund of the Korean Government and by the Korean Local Governments -- Gyeong\-sang\-buk-do Province and Pohang City -- and by the National Research Foundation of Korea (NRF) funded by the Korean government (MSIT) (grant number 2021R1A2C1010834).

\bibliography{main.bib}

\end{document}